\documentclass[aps,prl,preprint,superscriptaddress,showpacs,showkeys,11pt,a4paper]{revtex4-1} 
\usepackage{amsfonts}
\usepackage{amssymb}
\usepackage{amsmath}
\usepackage{graphicx}
\usepackage{color}

\begin{document}
\title{Universality and the collapse of multifractality in Barkhausen avalanches}

\author{Gustavo Zampier dos Santos Lima} 
\affiliation{Escola de Ci\^{e}ncias e Tecnologia, Universidade Federal do Rio Grande do Norte, 59078-970 Natal, RN, Brazil} 
\affiliation{Departamento de Biof\'{i}sica e Farmacologia, Universidade Federal do Rio Grande do Norte, 59078-970 Natal, RN, Brazil}
\affiliation{Keck Laboratory for Network Physiology, Department of Physics, Boston University, Boston, MA 02215, USA}
\author{Gilberto Corso}
\affiliation{Departamento de Biof\'{i}sica e Farmacologia, Universidade Federal do Rio Grande do Norte, 59078-970 Natal, RN, Brazil}
\author{Marcio Assolin Corr\^{e}a}
\affiliation{Departamento de F\'{i}sica, Universidade Federal do Rio Grande do Norte, 59078-900 Natal, RN, Brazil}
\author{Rubem Luis Sommer} 
\affiliation{Centro Brasileiro de Pesquisas F\'{i}sicas, Rua Dr.\ Xavier Sigaud 150, Urca, 22290-180 Rio de Janeiro, RJ, Brazil} 
\author{Plamen Ch.\ Ivanov}
\email[Electronic address: ]{plamen@buphy.bu.edu}
\affiliation{Keck Laboratory for Network Physiology, Department of Physics, Boston University, Boston, MA 02215, USA}
\affiliation{Harvard Medical School and Division of Sleep Medicine,
Brigham and Women Hospital, Boston, MA 02115, USA}
\affiliation{Institute of Solid State Physics, Bulgarian Academy of Sciences,
Sofia 1784, Bulgaria}
\author{Felipe Bohn}
\email[Electronic address: ]{felipebohn@fisica.ufrn.br}
\affiliation{Departamento de F\'{i}sica, Universidade Federal do Rio Grande do Norte, 59078-900 Natal, RN, Brazil}

\date{\today}



\maketitle 

\vspace{.5cm}

{\bf Barkhausen effect in ferromagnetic materials provides an excellent area for investigating scaling phenomena found in disordered systems exhibiting crackling noise~\cite{N410p242}. The critical dynamics is characterized by random pulses or avalanches with scale-invariant properties, power-law distributions, and universal features~\cite{N410p242,NP1p13,NP3p518,NC4p2927}. However, the traditional Barkhausen avalanches statistics may not be sufficient to fully characterize the complex temporal correlation of the magnetic domain walls dynamics. 
Here we go beyond power laws and focus on the multifractal scenario to quantify the temporal scaling characteristics of Barkhausen avalanches in polycrystalline and amorphous ferromagnetic films with thicknesses from $50$~nm to $1000$~nm. 
We show that the multifractal properties are dependent on the film thickness, and insensitive to the structural character of the materials. 
Further, we observe for the first time the collapse of the multifractality in the domain walls dynamics as the thickness is reduced, and the multifractal behavior gives place to a monofractal one over the entire range of time scales. 
The reorganization in the temporal scaling characteristics of Barkhausen avalanches is understood as an universal restructuring associated to the dimensional transition, from three to two-dimensional magnetization dynamics.}

Universal power laws are a central focus of investigations in statistical mechanics as they relate to underlying criticality in a wide variety of phenomena~\cite{Resumao_francesca}, such as plastic deformation and microfractures, shear response of a granular media, seismic activity in earthquakes, vortices dynamics in supercondutors, fluctuations in the stock market, and Barkhausen effect in ferromagnetic materials. 
Scaling and universality are associated with global organization of the system with long correlation and time scales that emerges in phase transitions close to the critical point. 
In ferromagnetic materials in the presence of an external magnetic field, the complex microscopic magnetization process and jerky motion of magnetic domain walls (DWs) are a classic example of self-organization and non-equilibrium critical dynamics~\cite{NP3p518}. 
The response to a smooth, slow external magnetic field is a dynamics exhibiting series of abrupt and irregular Barkhausen avalanches with a broad range of sizes and durations, that are characterized by universal scaling laws with critical exponents independent of the material microstructure~\cite{Resumao_BN}. 
Much efforts to understand the criticality in ferromagnetic materials have focused on the traditional probability distributions of Barkhausen avalanche sizes and avalanche durations, average avalanche size as a function of its duration, and average temporal avalanche shape~\cite{Resumao_francesca,Resumao_BN}.
Empirical investigations confronted to theoretical predictions and simulations have uncovered that critical exponents associated with such Barkhausen avalanches statistical properties reflect general features of the underlying magnetization dynamics~\cite{Resumao_francesca,Resumao_BN}.
Further, different scaling exponents have been verified according to the structural character of the sample, placing polycrystalline and amorphous materials in distinct universality classes associated with the range of interactions governing the DWs dynamics, as well as the exponents have been found to be dependent on the sample thickness, which is directly related to system dimensionality, thus splitting bulk material/thick films and thin films in distinct classes~\cite{Resumao_francesca,Resumao_BN, PRL84p4705, NP3p547, PRL90p0872031, NP7p316, PRE88p032811, PRE90p032821, PRL117p087201}.
However, the quantitative understanding of Barkhausen avalanches and DWs dynamics in ferromagnetic materials is far from complete. 
In contrast to this traditional Barkhausen statistics, investigations on the temporal structure of consecutive Barkhausen avalanches are still in the beginning~\cite{PRE86p066117} --- an analysis widely performed for several natural complex systems, as heartbeat dynamics~\cite{N399p461,C11p641}, earthquakes~\cite{PRE84p066123}, brain dynamics during sleep~\cite{EPL57p625}, fluctuations in financial markets~\cite{PRE78p036108}, and complex networks~\cite{PRE84p036118}.
Thus, doubts whether linear and nonlinear features of the temporal organization of avalanches are characterized by scaling laws that exhibit universality classes for different materials and whether temporal characteristics change with the thickness are questions that still remained elusive.

Here, we report an experimental investigation of the temporal scaling characteristics of Barkhausen avalanches in ferromagnetic films. 
Specifically, we ask whether the time series of avalanches caused by the irregular and irreversible motion of domain walls exhibit correlation properties characterized by monofractal and/or multifractal scaling features. Further, we address the question of if these features change with the universality class, i.e.\ they are influenced by the film dimensionality and range of interactions governing the DWs dynamics. 

We systematically analyze the temporal characteristics, over a wide range of time scales, of Barkhausen avalanches in polycrystalline NiFe and amorphous FeSiB ferromagnetic films with thicknesses from $50$~nm to $1000$~nm. 
The films split into three well-established classes of materials: (\textit{i}) Polycrystalline NiFe films thicker than $100$~nm, with universal three-dimensional magnetization dynamics governed by long-range dipolar interactions~\cite{NP7p316, PRE88p032811, PRL117p087201}; (\textit{ii}) Amorphous FeSiB films thicker than $100$~nm, with three-dimensional dynamics governed by short-range elastic interactions of the DWs~\cite{PRE90p032821, PRL117p087201}; (\textit{iii}) Both polycrystalline NiFe and amorphous FeSiB films with thicknesses below $100$~nm, presenting two-dimensional magnetization dynamics dominated by strong long-range dipolar interactions due to the appearance of DWs with zigzag morphology~\cite{NP3p547, PRL90p0872031,PRB89p104402} (see Methods section for details on the investigated films and experiment). 
Thus, the influence on the temporal characteristics of Barkhausen avalanches of the system dimensionality and range of interactions governing the DWs dynamics can be investigated in an experimentally controlled manner.

\begin{figure*}[!h]
\begin{center}
\includegraphics[width=17.0cm]{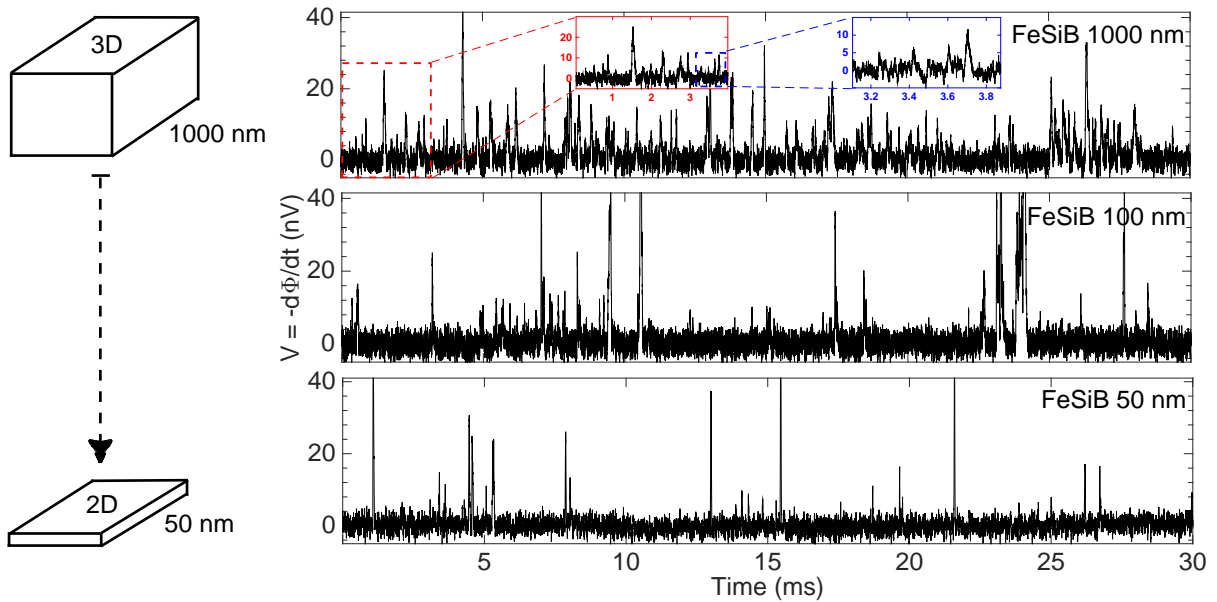}
\end{center}
\vspace{-.75cm}\caption{{\bf Representative Barkhausen noise in ferromagnetic films and its dependence with thickness.} 
In the left side, schematic representation of the dimensionality of the films with thickness. 
It is important to point out that films thicker than $100$~nm present three-dimensional magnetic behavior, while the ones thinner than $100$~nm are in the two-dimensional regime. 
In the right, experimental Barkhausen time series measured for amorphous FeSiB films with selected thicknesses. 
Similar behavior with thickness is verified for the polycrystalline NiFe films. 
Notice that the time series exhibit markedly different temporal characteristics with the dimensional transition occurring as the thickness is reduced from $100$~nm to $50$~nm. 
For the thicker films, above $100$~nm, the Barkhausen noise is inhomogeneous, with high amplitude and frequent avalanches associated with changes in the magnetization. 
In contrast, for thinner films, below $100$~nm, the noise is more homogeneous, with reduced occurrence of large avalanches. 
For the three-dimensional film with thickness of $100$~nm, the noise has intermediate characteristics compared to those observed for three-dimensional thicker and two-dimensional thinner films, suggesting the imminence of a dimensional transition. 
Such characteristics of the Barkhausen noise for films with different thicknesses connote distinct temporal organizations, implying dissimilar DWs dynamics. 
Barkhausen noise in different time windows of observation (insets in the top panel) reveals statistical self-similarity of avalanches at smaller scales, indicating an underlying scale-invariant temporal organization.}
\label{Fig_01}
\end{figure*}

Figure~\ref{Fig_01} shows representative experimental Barkhausen time series measured in ferromagnetic films with different thicknesses. 
Barkhausen noise is composed of discrete and irregular avalanches caused by the complex, jerky motion of the DWs during the magnetization process, and is related to sudden and irreversible changes in the magnetization. 
Remarkably, empirical observation of Barkhausen noise at different time scales reveal seemingly self-similar cascades formed by large, intermediate and small avalanches at each time window suggesting the presence of scale-invariant structure embedded in its temporal organization (see insets in the top panel in Fig.~\ref{Fig_01}). 
Further, the profile of Barkhausen noise significantly changes with the film thickness, i.e. system dimensionality, although it seems to be insensitive to the range of interactions governing the DWs dynamics, since similar behavior with thickness is verified for amorphous FeSiB and polycrystalline NiFe films (the latter not shown here). 
The Barkhausen noise for three-dimensional films thicker than $100$~nm is inhomogeneous, with high amplitude and frequent avalanches associated with changes in the magnetization through DWs motion. 
In contrast, the noise for two-dimensional ones, with thickness below $100$~nm, is more homogeneous, with reduced occurrence of large avalanches. 
Such characteristics of the Barkhausen noise for films with different thicknesses rise the hypothesis that the system dimensionality may be associated with distinct correlations and scaling temporal organization of Barkhausen noise, suggesting dissimilar underlying DWs dynamics. 

To probe for scale-invariant structure in the temporal organization of Barkhausen avalanches, and whether this structure changes with the system dimensionality and range of interactions governing the DWs dynamics, we apply the Detrended Fluctuation Analysis (DFA) method (see Methods section for details on the analysis). 

\begin{figure*}[!h]
\begin{center}
\includegraphics[width=17cm]{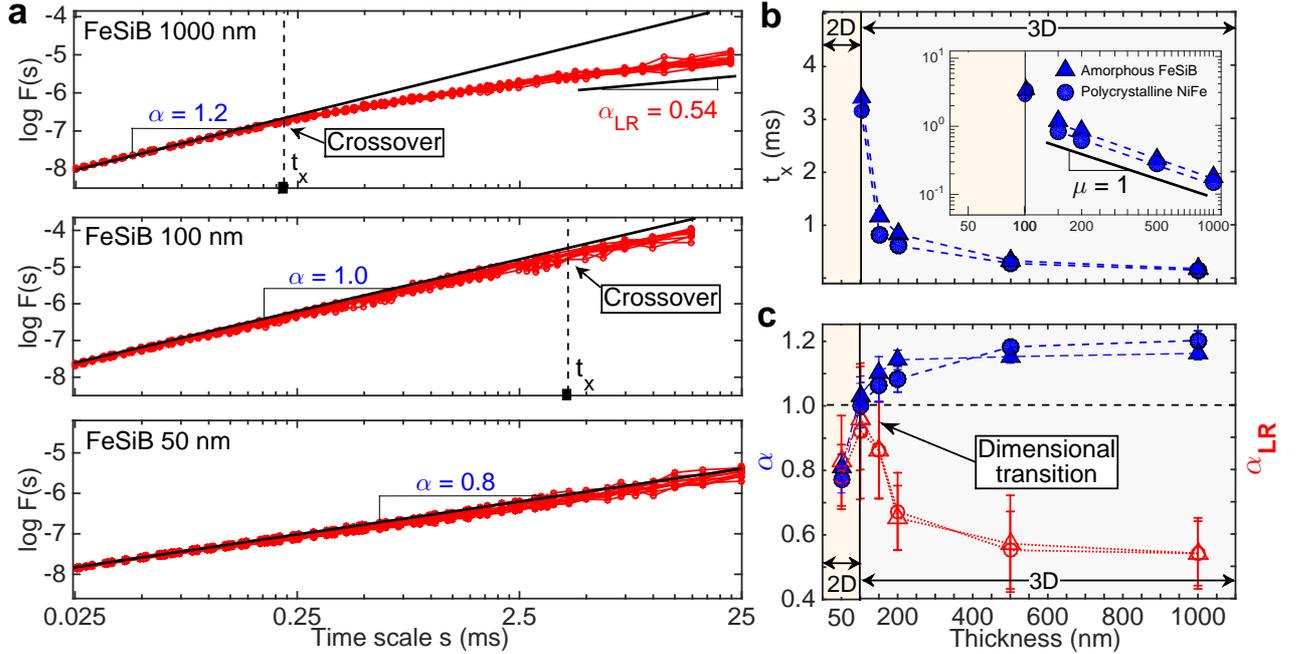}
\end{center}
\vspace{-.5cm}\caption{{\bf Dimensionality and transition in the temporal scaling characteristics of Barkhausen avalanches.} Barkhausen avalanches have different temporal characteristics in three and two-dimensional films. A peculiar behavior of the temporal scaling characteristics is observed at thickness of $100$~nm, characterizing an imminent transition. 
{\bf a}, Log-log plot of the fluctuation function $F(s)$ as a function of the time scale $s$, obtained from the DFA method (see Methods) applied to experimental Barkhausen noise time series measured in amorphous FeSiB films with the selected thicknesses of $1000$~nm, $100$~nm, and $50$~nm. 
Each panel in {\bf a} shows ten $F(s)$ curves, from different time series, indicating consistent scaling behavior. 
To guide the eyes, the black solid lines are power laws with the slope $\alpha$. 
{\bf b}, Dependence of the crossover time scale $t_\textrm{x}$ with thickness for amorphous FeSiB (blue triangles) and polycrystalline NiFe (blue circles) films. 
In the inset, $t_\textrm{x}$ as a function of the film thickness in the log-log scale, indicating a power-law scaling behavior. 
{\bf c}, Dependence of the scaling exponents $\alpha$ and $\alpha_{\textrm{LR}}$ with thickness for the amorphous FeSiB and polycrystalline NiFe films. 
Here, $\alpha$ (blue filled symbols) represents the correlation behavior at scales shorter than $t_\textrm{x}$, while $\alpha_{\textrm{LR}}$ (red open symbols) is obtained over long-range scales (for instance, above $3$~ms, as marked by the fitting line in the top panel of {\bf a}). 
In particular, for each film, the analyses are obtained from $200$ experimental Barkhausen time series, each of length of $2 \times 10^{5}$ points, corresponding to $50$~ms of duration. The error bars are estimated using the standard deviation.}
\label{Fig_02}
\end{figure*}

Figure~\ref{Fig_02} shows the general behavior of the fluctuation function with the film thickness. 
Amorphous FeSiB and polycrystalline NiFe films share similar behavior (the latter not shown here).
For all thicknesses, the Barkhausen noise exhibits correlations of a power-law type, indicating a robust scale-invariant organization of the avalanches over a broad range of time scales. 
However, the scaling exponent associated with this scale-invariant behavior significantly changes with the thickness. 
For the $1000$~nm-thick film (top panel in Fig.~\ref{Fig_02}a), we clearly see a pronounced crossover at the time scale $t_\textrm{x} \approx 0.25$~ms, from a regime with robust power-law correlations spanning over one decade at short time scales, characterized by a scaling exponent $\alpha = 1.20 \pm 0.02$, to a regime close to random behavior with an exponent $\alpha_{\textrm{LR}} = 0.54 \pm 0.05$ for large time scales. 
We note that this crossover to random behavior at large time scale is not an artifact of electronic white noise present in the experimental recordings. 
The amplitude of electronic noise is one magnitude smaller than Barkhausen noise, therefore the contribution of electronic noise to the fluctuation function $F(s)$ is negligible. 
Moreover, external white noise does affect the scaling of long-range positively correlated fractal signals, but only at very short time scales and when the standard deviation of the white noise exceeds by a decade the amplitude of fluctuations in the correlated signal~\cite{PRE64p011114}.
In this way, for three-dimensional films, our results indicate a genuine crossover in the Barkhausen avalanches temporal characteristics, from strongly correlated (at short-time scales) to random (at long-time scales). 

With decreasing thickness we observe that the crossover at $t_\textrm{x}$ gradually shifts to longer time scales. For the $50$~nm-thick film, with two-dimensional magnetic behavior, we find that the temporal dynamics of Barkhausen avalanches exhibits scale-invariant behavior with long-range power-law correlations and exponent $\alpha = 0.80 \pm 0.04$ over the entire range of the time scales, spanning more than three decades (bottom panel in Fig.~\ref{Fig_02}a).
Thus, we verify that Barkhausen avalanches have markedly different temporal characteristics in films with different dimensionality, i.e.\ a typical crossover from correlated to uncorrelated behavior for films with three-dimensional magnetic behavior, and in contrast, absence of crossover and a single scaling behavior in two-dimensional films. 
Remarkably, we find that for the $100$~nm-thick film (middle panel in Fig.~\ref{Fig_02}a), practically over the entire range of time scales, a scaling behavior with a critical exponent $\alpha =1.00 \pm 0.03$ emerges, indicating an imminent transition in the temporal scaling characteristics of the avalanches at the border between three and two-dimensional behaviors. 

Further, we investigate the dependence of the crossover time scale $t_\textrm{x}$ and the scaling exponents $\alpha$ and $\alpha_{\textrm{LR}}$ with the thickness. Our analyses show that $t_\textrm{x}$ remains relatively stable over a broad range of thicknesses, from $1000$~nm to $200$~nm, abruptly increases with the approach of the thickness to $100$~nm, and becomes infinite (i.e.\ reaches the finite size of the recorded time series) for films with thickness below $100$~nm (Fig.~\ref{Fig_02}b). 
In particular, we verify that the thickness dependence of $t_\textrm{x}$ follows a power-law scaling behavior with slope $\mu=1.0$ (see inset of Fig.~\ref{Fig_02}b).
We observe a simultaneous evolution for the scaling exponents $\alpha$ and $\alpha_{\textrm{LR}}$. The $\alpha$ value, which characterize the DWs dynamics at short time scales, remains $\approx 1.2$ for thicknesses in the broad interval $1000$~nm to $200$~nm, indicating strong avalanche correlations for time scales below $t_\textrm{x}$, abruptly decreases to a critical value of $\alpha \approx 1.0$ at $100$~nm, and drops to $\alpha \approx 0.8$ over the entire range of the time scales for films thinner than $100$~nm (Fig.~\ref{Fig_02}c). 
In contrast to $\alpha$, the scaling exponent $\alpha_{\textrm{LR}}$, characterizing DWs dynamics at large time scales, is $\approx 0.54$ for thicknesses in the interval between $1000$~nm to $200$~nm, indicating an uncorrelated behavior, abruptly increases converging to $\alpha_{\textrm{LR}} = \alpha \approx 1.0$ at $100$~nm, and coincides with $\alpha$ for thickness below $100$~nm. 
It is important to notice that in the border line at thickness $\approx 100$~nm, the peculiar behavior of the scaling characteristics $t_\textrm{x}$, $\alpha$ and $\alpha_{\textrm{LR}}$ occurs simultaneously to the dimensional transition in the magnetic behavior, from three to two-dimensional magnetization dynamics. 
Notably, the scaling behaviors are consistent for all Barkhausen noise time series recorded for each film, as evidenced by the small error bars for the scaling exponents and by the statistical significance tests (Fig.~\ref{Fig_02}c). 
Remarkably, we also find that both amorphous FeSiB and polycrystalline NiFe films exhibit the very same scaling behavior with the same exponent and the same dependence with thickness.

In the sequence, we test whether the Barkhausen avalanches exhibit multifractal behavior, an analysis that requires a spectrum of exponents, i.e.\ a multifractal spectrum, to fully characterize the temporal structure. 
We perform a scaling analysis of the fluctuation function $F_q(s)$ for a range of $q$ moments applying the Multifractal Detrended Fluctuation Analysis (MF-DFA) method (see Methods section for details on the analysis). 

\begin{figure*}[!h]
\begin{center}
\includegraphics[width=17cm]{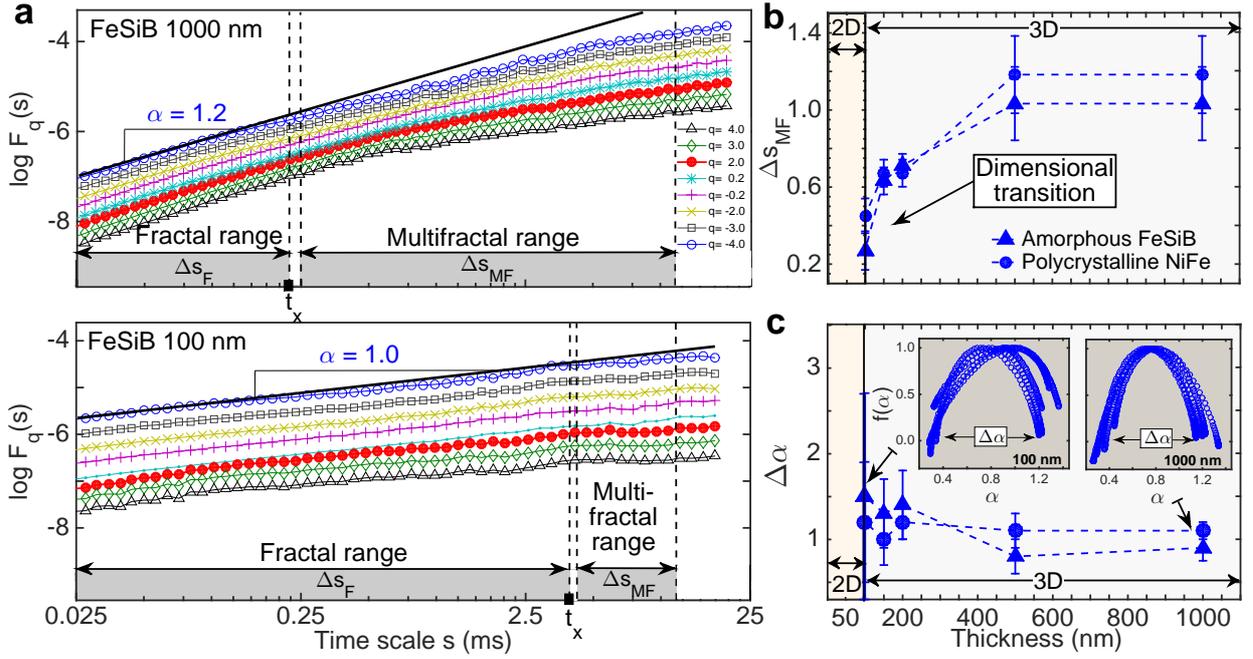}  
\end{center}
\vspace{-.5cm}\caption{{\bf Multifractality and multifractal collapse in Barkhausen avalanches.} 
Multifractal behavior of Barkhausen avalanches is strongly dependent on the dimensionality. As the thickness is reduced from $100$~nm to $50$~nm, the multifractality of the DWs dynamics collapses, and the multifractal behavior gives place to a monofractal one over the entire range of time scales. 
{\bf a}, Log-log plot of the fluctuation function $F_{q}(s)$ as a function of the time scale $s$ for several moments $q$, ranging from $q=-4.0$ to $q=+4.0$, obtained from the MF-DFA method (see Methods), for the amorphous FeSiB films with the selected thicknesses of $1000$ and $100$~nm. 
In particular, these scaling curves represent the analysis of a single Barkhausen noise time series, and the scaling curve for $q=2$, corresponding to the DFA fluctuation function $F(s)$, and $t_\textrm{x}$, the crossover time scale, are the very same previously presented in Fig.~\ref{Fig_02}. 
To guide the eyes, the black solid lines are power laws with slope $\alpha$. 
The temporal dynamics of Barkhausen avalanches is characterized by two temporal scaling regimes: fractal range $\Delta s_{\textrm{F}}$ and multifractal range $\Delta s_{\textrm{MF}}$, which are separated by the crossover $t_\textrm{x}$.
{\bf b}, Range of time scales $\Delta s_{\textrm{MF}}$, in units of decades, in which the Barkhausen avalanches exhibit multifractal behavior, as a function of the thickness. 
{\bf c}, The width $\Delta \alpha$ of the multifractal spectrum $f(\alpha)$ as a function of film thickness. 
The insets show the multifractal spectrum $f(\alpha)$ of five representative Barkhausen noise time series for the amorphous FeSiB films with thicknesses of $100$~nm and $1000$~nm. 
In particular, for the $50$~nm-thick film, $\Delta s_{\textrm{MF}}$ and $\Delta \alpha$ are not shown in {\bf b} and {\bf c} since the multifractality collapses for thickness just below $100$~nm. 
In particular, for each film, the analyses are obtained from $200$ experimental Barkhausen time series, each of length of $2 \times 10^{5}$ points, corresponding to $50$~ms of duration. The error bars are estimated using the standard deviation.
}
\label{Fig_03}
\end{figure*}

Figure~\ref{Fig_03} shows the behavior of the generalized $q$ dependent fluctuation function with the film thickness.
Remarkably, Amorphous FeSiB and polycrystalline NiFe films share similar behavior (the latter not shown here). 
For the $1000$~nm-thick film we find that the dynamics at short time scales $\Delta s_{\textrm{F}}$ exhibits the same scaling behavior shown in Fig.~\ref{Fig_02}, with exponent $\alpha \approx 1.2$ for all $q$ moments, indicating a monofractal behavior. In addition, at intermediate time scales $\Delta s_{\textrm{MF}}$ the fluctuation function $F_q(s)$ is characterized by a distinct scaling exponents for different $q$ moments, suggesting a multifractal behavior (top panel in Fig.~\ref{Fig_03}a). 

A crossover from monofractal to multifractal behavior is consistently verified for the films with three-dimensional magnetic behavior, irrespective on the thickness. 
However, with decreasing the thickness, the crossover $t_\textrm{x}$ shifts, and the range of time scales $\Delta s_{\textrm{F}}$, corresponding to monofractal scales, expands from short to intermediate and long time scales. 
At the same time, the range of scales $\Delta s_{\textrm{MF}}$, where the multifractality is observed, abruptly shrinks (bottom panel in Fig.~\ref{Fig_03}a). This can be clearly visualized in  Fig.~\ref{Fig_03}b, which shows the evolution of the estimation of the multifractal time scales $\Delta s_{\textrm{MF}}$ with thickness. 
For the $100$~nm-thick film, the Barkhausen avalanches exhibit a monofractal behavior with exponent $\alpha \approx 1.0$ roughly spanning the entire time scales range, and the dynamics exhibits multifractality only in a very short range $\Delta s_{\textrm{MF}} < 0.5$ decades (bottom panel in Fig.~\ref{Fig_03}a). 
For the thinner film with $50$~nm the multifractality collapse (the time scales multifractal range $\Delta s_{\textrm{MF}}$ vanishes) and Barkhausen avalanches exhibit monofractal behavior over the entire range of time scales characterized by exponent $\alpha \approx 0.8$ for all $q$ moments. 

The results indicate that the three-dimensional films exhibit DWs dynamics with high degree of multifractality.
Figure~\ref{Fig_03}c shows that the robust multifractal spectrum with magnitude $\Delta \alpha \approx 1.0$ remains stable for the thickness range $1000 $~nm to $100$~nm, although $\Delta s_{\textrm{MF}}$ significantly changes with the proximity to $100$~nm.
Despite the $\Delta \alpha$ stability, the exponents error bar increases as the thickness approaches $100$~nm, a fingerprint associated to the imminent vanishing of range of time scales $\Delta s_{\textrm{MF}}$.
Thus the multifractality gives place to a monofractal behavior in the two-dimensional regime, films with thickness of $50$~nm, with the transition occurring just below $100$~nm. 

Our findings raise an interesting issue about the DWs dynamics in ferromagnetic films. 
The analyses of Barkhausen avalanches show statistically similar results for the temporal scaling characteristics for both amorphous and polycrystalline ferromagnetic materials, placing them in a single universality class. 
Moreover, we provide experimental evidence that the temporal scaling characteristics of Barkhausen avalanches are strongly influenced by the film dimensionality, but are insensitive to the range of the interactions governing the DWs dynamics.
We interpret the reorganization in temporal scaling characteristics of Barkhausen avalanches as an indicator of an universal restructuring associated to the dimensional transition of the magnetic behavior occurring as the thickness is reduced, from a three-dimensional DWs dynamics in films thicker than $100$ nm~\cite{NP7p316, PRE88p032811, PRE90p032821, PRL117p087201} to a two-dimensional regime, commonly verified for films thinner than $50$~nm~\cite{NP3p547, PRL90p0872031}. 

Much of the critical behavior observed in nature can be explained by range of interactions and system dimensionality. 
On the other hand, the understanding of the influence that these general properties have on the monofractal/multifractal behavior is far from complete. 
Our experiments provide evidence that the multifractal properties are dependent on the film thickness, and insensitive to the structural character of the materials. 
In fact, we understand that the strong multifractal behavior is due to the mixing of several correlated processes with distinct temporal correlations lengths, the hypothesis of overlapping of several non-correlated avalanches. 
As the thickness is reduced from $100$~nm to $50$~nm regardless the structural character of the films, the multifractality of the DWs dynamics collapses, and the multifractal behavior gives place to a monofractal one over the entire range of time scales.
The reorganization in the temporal scaling characteristics of Barkhausen avalanches is understood as an universal restructuring associated to a dimensional transition of the magnetic behavior, from a three to a two-dimensional magnetization dynamics, occurring within this thickness range.
In this sense, in the limit of the dimensional transition, the magnetic energy that occupied the bulk space is squeezed to the plane, increasing the coupling among magnetic domains. 
The correlation of the time series reflects the stronger magnetic coupling in two-dimensional films. 
Increasing the correlation, the fractal regime becomes unique and robust and, as a consequence, the multifractality vanishes. 
Thus, our work also demonstrates that the multifractal analysis is a powerful tool to investigate systems exhibiting crackling noise and appears as a sharper test going beyond power laws. 
We are confident that our results trigger interesting challenges to theoreticians and further experimental investigations in diverse systems exhibiting crackling noise.

\section*{Methods}

{\bf Ferromagnetic films and experiment.} We perform Barkhausen experiments in polycrystalline Ni$_{81}$Fe$_{19}$ (NiFe) and amorphous Fe$_{75}$Si$_{15}$B$_{10}$ (FeSiB) ferromagnetic films with the thicknesses of $50$, $100$, $150$, $200$, $500$, and $1000$~nm. 
The films are deposited by magnetron sputtering onto glass substrates, with dimensions $10$ mm $\times$ $4$ mm, covered with a $2$~nm thick Ta buffer layer. 
The deposition process is carried out with the following parameters: base vacuum of $10^{-7}$ Torr, deposition pressure of $5.2$ mTorr with a $99.99$\% pure Ar at $20$ sccm constant flow, and DC source with current of $50$~mA and $65$~W set in the RF power supply for the deposition of the Ta and ferromagnetic layers, respectively. 
During the deposition, the substrate moves at constant speed through the plasma to improve the film uniformity, and a constant magnetic field of $1$ kOe is applied along the main axis of the substrate in order to induce magnetic anisotropy. 
X-ray diffraction is employed to calibrate the sample thicknesses and also to verify the structural character of all films. 
Quasi-static magnetization curves are obtained along and perpendicular to the main axis of the films, in order to verify the magnetic properties. 
Detailed information on the structural and magnetic characterizations is found in Refs.~\cite{NP7p316,PRE88p032811,PRE90p032821}. 

We record Barkhausen noise using the traditional inductive technique in an open magnetic circuit, in which one detects time series of voltage pulses with a sensing coil wound around a ferromagnetic material submitted to a slow-varying magnetic field. 
In our setup, sample and pick up coils are inserted in a long solenoid with compensation for the borders, to ensure an homogeneous magnetic field on the sample. 
The sample is driven by a triangular magnetic field, applied along the main axis of the sample, with an amplitude high enough to saturate it magnetically. 
Barkhausen noise is detected by a sensing coil wound around the central part of the sample. 
A second pickup coil, with the same cross section and number of turns, is adapted in order to compensate the signal induced by the magnetizing field. 
The Barkhausen signal is then amplified, filtered, and finally digitalized. 
Barkhausen noise measurements are performed using a sensing coil with $400$ turns, $3.5$ mm long and $4.5$ mm wide, and under similar conditions, i.e.\ $50$~mHz triangular magnetic field, $100$ kHz low-pass filter set in the preamplifier and signal acquisition with sampling rate of $4\times 10^6$ samples per second. 
The time series are acquired just around the central part of the hysteresis loop, near the coercive field, where the domain walls motion is the main magnetization mechanism~\cite{Resumao_BN}. 
In particular, for each ferromagnetic film, the following analyses are obtained from $200$ time series, each one with length of $2 \times 10^{5}$ points, corresponding to $50$~ms of duration.

The universality class of the Barkhausen noise in a sample is commonly identified by measuring the distributions of Barkhausen avalanche sizes and avalanche durations, the average avalanche size as a function of its duration, and the average temporal avalanche shape~\cite{Resumao_BN}. 
The first three statistical functions typically display scaling in a quite large range and are respectively described by the critical exponents $\tau$, $\alpha'$, and $1/\sigma\nu z$~\footnote{To avoid misunderstanding with the used notation, it is important to keep in mind that, in the traditional Barkhausen noise statistical analysis, we assume $\tau$ and $\alpha'$ as the critical exponents measured from the distributions of avalanche sizes and avalanche durations, respectively. 
On the other hand, in the context of the multifractal analysis, the similar symbols $\tau(q)$ and $\alpha$ present distinct meanings, corresponding to the multifractal scaling exponent and H{\"o}lder exponent, respectively.}, as well as the average shape evolves with the universality class, and corroborates the exponent $1/\sigma\nu z$~\cite{Resumao_BN,NC4p2927}. 

Several theoretical models have been proposed to explain the DWs dynamics and the traditional Barkhausen noise statistical properties. 
Taking into account theoretical predictions found in literature~\cite{PRL79p4669,PRB89p104402,PRE69p026126}, we interpret our experimental data in terms of different universality classes, according to the system dimensionality and range of interactions governing the DWs dynamics. 
Our experiments allow us to infer that the exponents measured for the polycrystalline NiFe and amorphous FeSiB films with distinct thicknesses assume values consistent with three well-defined universality classes. 
Thus, we analyse the temporal characteristics of Barkhausen avalanches in the following classes of materials: (\textit{i}) Polycrystalline NiFe films with thicknesses above $100$~nm, characterized by exponents $\tau \sim 1.5$, $\alpha'\sim 2.0$, and $1/\sigma \nu z \sim 2.0$, with universal three-dimensional magnetization dynamics governed by long-range dipolar interactions~\cite{NP7p316, PRE88p032811, PRL117p087201}; (\textit{ii}) Amorphous FeSiB films with thicknesses above $100$~nm with exponents $\tau \sim 1.27$, $\alpha' \sim 1.5$, and $1/\sigma \nu z \sim 1.77$, presenting three-dimensional dynamics governed by short-range elastic interactions of the DWs~\cite{PRE90p032821, PRL117p087201}; (\textit{iii}) Polycrystalline NiFe and amorphous FeSiB films with thicknesses below $100$~nm, with $\tau \sim 1.33$, $\alpha' \sim 1.5$, and $1/\sigma \nu z \sim \vartheta \sim 1.55$, indicating two-dimensional magnetization dynamics dominated by strong long-range dipolar interactions due to the appearance of DWs with zigzag morphology~\cite{NP3p547, PRL90p0872031,PRB89p104402}.

{\bf Data Analysis.} We employ the Detrended Fluctuation Analysis (DFA) (presented in Fig.~\ref{Fig_02}) and the  generalized Multifractal Detrended Fluctuation Analysis (MF-DFA) (shown in Fig.~\ref{Fig_03}) methods~\cite{PA316p87,FP3p141,PRE86p066117} 
to investigate the temporal nonlinear characteristics of Barkhausen avalanches in ferromagnetic films. 

The Detrended Fluctuation Analysis (DFA) method~\cite{PRE49p1685} been developed initially to quantify dynamic characteristics of physiological fluctuations embedded in nonstationary physiological signals. 
Compared with traditional correlation analyses, such as autocorrelation, power spectrum, and Hurst analysis, the advantage of the DFA method resides in the accurate quantification of the correlation property of signals masked by polynomial trends~\cite{PRE64p011114}. 
The DFA method quantifies the detrended fluctuation function $F(s)$ of a signal at different time scales $s$. 
A power-law functional form $F(s) \approx s^{\alpha}$ indicates the presence of self-similar fractal organization in the fluctuations. The parameter $\alpha$, here also called scaling exponent, quantifies the correlation properties of the signal. 
Furthermore, $\alpha = 0.5$ indicates signal absence of correlations, similarly to a white noise; if $\alpha = 1.5$, the signal behaves as a random walk, and indicates a Brownian motion-like dynamics; if $ 0.5 < \alpha < 1.5 $, there are positive correlations, i.e.\ large avalanche intervals are more likely to be followed by large intervals (and vice versa for small avalanche intervals); and if $\alpha < 0.5$, the signal is anticorrelated, i.e.\ large avalanche intervals are likely to be followed by small intervals (vice versa for small avalanche intervals), with stronger anticorrelations when $\alpha$ is closer to $0$. 
A striking advantage of the DFA method resides in the fact that it allows the quantification of signals with $\alpha > 1.0$, which cannot be done using the traditional autocorrelation and $R/S$ analyses, as well as of signals with strong anticorrelations~\cite{PRE64p011114}. 
In contrast to the conventional methods, the DFA one avoids spurious detection of apparent long-range correlations that are artifacts of nonstationarity~\cite{PRE71p011104}. Thus, the DFA method is able to detect subtle temporal structures in highly heterogeneous time series. 
However, an inherent limitation of the DFA analysis is the maximum time scale $s_{max}$ for which the fluctuation function $F(s)$ can be reliably calculated. 
To overcome this issue and ensure sufficient statistics at large scales, $s_{max}$ shall be chosen $s_{max} \leq N/6$, where $N$ is the length of the signal~\cite{PRE64p011114}.

The DFA method quantifies linear fractal characteristics related to two-point correlation, while the MF-DFA one probes long-term nonlinear properties of the time series. 
In particular, it has been already verified that signals with identical self-similar temporal organization, quantified by the DFA scaling exponent $\alpha$, can exhibit very different nonlinear properties captured by the MF-DFA method~\cite{IEEETBE56p1564}.

The generalized multifractal DFA method consists of a sequence of five steps, in which the first ones are essentially identical to that of the conventional DFA procedure. 
In this sense, the MF-DFA is an extension of the DFA method with a range of $q$, i.e.\ an average over all segments to obtain the $q$th-order fluctuation function $F_q(s)$ where, in general, the index variable $q$ can assume any real value, except zero. 

Initially we assume that $x_{k}$ is an experimental Barkhausen noise time series, of length $N$. 
Thus, in the first step, the accumulated profile from a time series, so-called random walk like time series~\cite{FP3p141}, is determined by the following equation
\begin{eqnarray}\label{in01}
Y(i) \equiv \sum_{k=1}^{i}{[x_k - \left\langle x \right\rangle]}, i = 1,..., N,
\end{eqnarray}
\noindent where $\left\langle x \right\rangle$ denotes the mean of the time series $x_k$. 

In step two, for a given time scale $s$, the accumulated profile from Eq.~(\ref{in01}) is divided into~$N_s \equiv \textrm{int}(N/s)$ integer disjoint segments of equal length $s$. 
In step three, for each one of the $N_s$ segments, the local trend is determined by a polynomial fitting of the data, and then the variance for each segment $\nu =  1,...,N_s$ is estimated through: 
\begin{eqnarray}\label{fluctuation}
F^{2}(\nu,s) \equiv \frac{1}{s}\sum_{i=1}^{s}\left\{Y[(\nu - 1)+i] - y_{\nu}(i)\right\}^2,
\end{eqnarray}
\noindent where $y_{\nu}(i)$ is the polynomial fitting for the segment $\nu$.

In step four, the average of the variances over all segments is computed to obtain the $q$th-order fluctuation function $F_q(s)$. In this case, for $q \neq 0$, the fluctuation function is given by: 
\begin{eqnarray}\label{Fq}
F_q(s) \equiv \left\{\frac{1}{N_s}\sum_{\nu=1}^{N_s}\left[ F^{2}(\nu,s) \right]^{q/2} \right\}^{1/q},
\end{eqnarray}
\noindent while for $q = 0$,
\begin{eqnarray}\label{Fqa}
F_0(s) \equiv \exp\left\{\frac{1}{N_s}\sum_{\nu=1}^{N_s}\ln\left[ F^{2}(\nu,s)
\right] \right\}.
\end{eqnarray}

In particular, for $q=2$, the standard DFA procedure is retrieved.
However, here, we are concerned with how the generalized $q$ dependent fluctuation function $F_q(s)$ depends on the time scale $s$, for some values of $q$. 
For this purpose, steps two to four must be repeated for different values of time scale $s$. 
According to Ref.~\cite{PA316p87}, for very large scales $s\geq N/4$, the employed procedure becomes statistically unreliable, due to the number of segments $N_s$ averaging in Eqs.~(\ref{Fq}) and (\ref{Fqa}) be very small. 
Thus, we take for our Barkhausen signal analysis the maximum scale value of $N/4$.

Finally, in the last step, the scaling behavior of the fluctuation functions $F_q(s)$ is estimated by the slope of the plot of $\log_{10}[F_q(s)]$ {\it vs.}\ $\log_{10}[s]$, for a range of $q$ values. 
In particular, we use $q$ between $-4$ to $4$. 
If the experimental Barkhausen noise time series present a long-range power-law correlation  $F_q(s)$ increases, for sufficiently large values of $s$, following the power-law scaling given by: 
\begin{eqnarray}\label{PowerLaw}
F_q(s) \approx s^{h(q)},
\end{eqnarray}
\noindent where $h(q)$ is the so-called generalized Hurst exponent. 

To estimate the $h(q)$   for several  $q$ we regress $h(q)$ on $F_q(s)$, Eq.~(\ref{PowerLaw}). 
Thereby, strengthening this idea, for monofractal time series,  $h(q)$ is independent of $q$, since variance scale behavior   $F^{2}(\nu,s)$ is identical for all segments $\nu$, resulting in $h(q)=H$. 
In opposition, when   small and large fluctuations differ  it  is observed   dependence of $h(q)$ over $q$ which 
characterizes the multifractal behavior.

From this point, the multifractal scaling exponent $\tau(q)$ can be determined from $h(q)$ by the relation
\begin{eqnarray}\label{tau_q}
\tau(q) = q \times h(q) - 1.
\end{eqnarray}
\noindent In this case, if there is a linear dependence of the spectrum $\tau(q)$ with $q$, the time series is considered monofractal, otherwise it is multifractal. 

Futhermore, it is possible to characterize the multifractality of time series by considering the multifractal spectrum $f(\alpha)$, where $\alpha$ is the H{\"o}lder exponent. 
The multifractal spectrum $f(\alpha)$ is related to $\tau(q)$ via a Legendre transform~\cite{Feder} 
\begin{eqnarray}
\alpha = \tau'(q),
\end{eqnarray}
\noindent and
\begin{eqnarray}\label{f_alpha}
f(\alpha) = q\alpha - \tau(q).
\end{eqnarray}

The magnitude of multifractality in time series can be determined by the width of the spectrum $\Delta\alpha = \alpha_{max} - \alpha_{min}$. 
Intuitively, the wider is the multifractal spectrum, the   stronger is the multifractality of the time series.

\section*{Acknowledgments}
~The research has been supported by the Brazilian agencies CNPq (Grants No.~$306423$/$2014$-$6$, No.~$471302$/$2013$-$9$, No.~$306362$/$2014$-$7$, and No.~$441760$/$2014$-$7$), CAPES (BEX $11264$/$13$-$6$), and FAPERN (Pronem No.~$03/2012$).

\section*{Author Contributions}
~F.B., M.A.C., and R.L.S. performed and analysed the experiments. G.Z.S.L., G.C., F.B. and P.C.I. performed the multifractal analysis. F.B., G.Z.S.L., G.C., and P.C.I. wrote the first draft of the manuscript. All authors contributed to improve the manuscript.

\section*{Competing Financial Interests statement}
~The authors declare no competing financial interests. 

\section*{Additional information}
~Correspondence and requests for materials should be addressed to P.C.I. and F.B.

\newpage
\section*{Figure Legends}

\newpage

\newpage

\end{document}